\documentstyle[seceq]{ptptex}

\def\med{{1\ov 2}}
\def\hepth#1{ {\tt hep-th/#1}}

\def\be{\begin{equation}}
\def\ee{\end{equation}}
\def\bes{\begin{equation*}}
\def\ees{\end{equation*}}

\def\beqa{\begin{eqnarray}}
\def\beqas{\begin{eqnarray*}}
\def\eeqa{\end{eqnarray}}
\def\eeqas{\end{eqnarray*}}
\def\bea{\begin{eqnarray}}
\def\eea{\end{eqnarray}}

\def\cl{\mbox{\tiny (class)}}

\def\H{{\cal H}}
\def\tr{{\rm tr}}

\def\F{{\cal F}}

\def\d{\partial}
\def\ov{\over}

\def\pder#1#2{{{\partial #1}\over{\partial #2}}}
\def\ppder#1#2#3{{\partial^2 #1\ov\partial #2\partial #3}}
\def\dpder#1#2{{\partial^2 #1\ov\partial #2 ^2 }}
\def\bemat{\left(\begin{array}}
\def\enmat{\end{array}\right)}
\def\theequation{\thesection.\arabic{equation}}

\def\Fppk{\vec \alpha\!\cdot\!\!\F''_k\!\!\cdot\! \vec \alpha}
\def\Fppu{\vec \alpha\!\cdot\!\!\F''_1\!\!\cdot\! \vec \alpha}

\title{The uses of Whitham hierarchies \footnote{Talk given at the
workshop ``Gauge theory and integrable models" (YITP, Kyoto), January 26-29, 1999.}
}

\author{ Marcos {\sc   Mari\~no}\footnote{marcos.marino@yale.edu} }

\inst{Department of Physics, Yale University\\ New Haven, CT 06520,
 USA.
}

\abst{We review some of the uses of Whitham hierarchies in the context of
the theory of the prepotential in ${\cal N}=2$ supersymmetric gauge
theories. We focus on the structure of the contact terms in the
twisted topological theory, and on the connection between Whitham
hierarchies and the $u$-plane integrals for higher rank gauge
groups, trying to put together the different approaches involved in
this connection. We also review two other uses of the Whitham
hierarchies: the interpretation of the slow times as supersymmetry
breaking parameters, and the new techniques to extract instanton
corrections using the RG equations written in terms of theta functions.}

\begin{document}

\maketitle

\section{Introduction}

Shortly after Seiberg and Witten
discovered the exact solution
for the low energy effective action of ${\cal N}=2$
supersymmetric $SU(2)$ gauge theories
\cite{rf:sw}, it was realized that
this solution, as well as the
generalizations to other gauge groups,
could be interpreted in the framework
of integrable systems \cite{rf:toda,rf:dw}.
This connection, which is largely based
on a common description in
terms of Riemann surfaces and their
Jacobians, has essentially involved two
ingredients: the first ingredient has been
the identification of an integrable classical
mechanical system whose associated spectral
curve reproduces the curve describing the low energy
dynamics of the gauge theory. For example,
for pure Yang-Mills theories, the
relevant integrable system turns out to
be the periodic Toda chain \cite{rf:todamar},
while in the case of the mass deformed
${\cal N}=4$ theory, the corresponding system
is Calogero-Moser \cite{rf:cm,rf:im,rf:dhokercm}
(which is in turn equivalent to the Hitchin
system described in \cite{rf:dw}). The second
ingredient is the theory of the
prepotential: once the integrable system has been identified, one
considers the quasiclassical Whitham hierarchy associated
to the original hierarchy, which is constructed
by introducing ``slow" times instead of the
original, ``fast" times \cite{rf:krich}. The prepotential of the effective
theory turns out to be, essentially, the logarithm of the quasiclassical
tau function and hence depends on the slow
times of the corresponding Whitham hierarchy \cite{rf:toda,rf:prepw,rf:im}.

Although there are general and rigorous arguments showing that
the effective ${\cal N}=2$ theories should be governed by
 integrable systems \cite{rf:dw,rf:freed}, there is for the moment
no dynamical reason to explain why they are
described by these particular one-dimensional mechanical systems. Nevertheless,
this remarkable connection between two {\it a priori}
unrelated fields has been very fruitful. For example, the connection to
integrable systems gives a unifying approach to find the Seiberg-Witten curves
for the different gauge groups, because both Toda and Calogero-Moser systems
can be formulated for any Lie algebra. In this paper, we will focus
on the second aspect of the connection, namely the theory of the prepotential
in the framework of Whitham hierarchies, and we will try to show that
this framework is not only an elegant way to describe the prepotential, but also
the appropriate point of view to understand some important
aspects of ${\cal N}=2$ gauge theories. The main reason for this is that within the
Whitham approach one can consider an ``enlarged" prepotential with extra parameters.
Consider, for simplicity, the case of pure $SU(N)$ super Yang-Mills theory. In this case,
the usual Seiberg-Witten prepotential ${\cal F} (a^I, \Lambda)$ is a holomorphic
function of the coordinates of
the moduli space, $a^I$, where $I=1, \cdots, N-1$,
and the dynamically generated scale of the theory, $\Lambda$. The theory of
Whitham hierarchies provides a generalized prepotential ${\cal F} (a^I, T_n)$,
where $T_n$, $n=1, \cdots,
N-1$, are the slow times of the Whitham hierarchy. This prepotential is a {\it deformation} of
the Seiberg-Witten prepotential, in the sense that, if we put $T_1=\Lambda$, $T_{n>1}=0$,
we recover the original function ${\cal F} (a^I, \Lambda)$.

There are two contexts where deformations of the prepotential can be relevant. The first
context is the study of contact terms in the twisted version of ${\cal N}=2$ gauge theories. These
contact terms, as we will explain, appear when one computes the generating functional of a
certain class of operators, and the source terms for insertions of these
operators can be regarded as deformations of the original action. The other context is the study
of soft supersymmetry breaking terms, which can also be understood as deformations of the theory.
In both cases, the deformations of the action can be described in terms of the Whitham time
variables of the enlarged prepotential. In this paper, we will mainly focus on the
relation between the contact terms in twisted
${\cal N}=2$ theories and the theory of the
prepotential from Whitham hierarchies. The issue of contact terms has been
previously addressed in \cite{rf:mw,rf:moore1,rf:moore2,rf:moore3,rf:lns,rf:taka,rf:ITEP}
from different points of view. We will try to put together
these approaches using the relation to integrable systems as a unifying framework.
We will also review the uses of Whitham hierarchies in the study of soft supersymmetry
breaking, as explained in \cite{rf:our}.

An extra bonus of the Whitham approach to the theory of the prepotential
is a set of renormalization group (RG) equations for the prepotential
in terms of theta functions associated to the root lattice of the
gauge group \cite{rf:ITEP}. It was shown in \cite{rf:our} that these equations
give an elegant way to derive the instanton expansion of the prepotential
following a recursive procedure. We will also review very briefly the
strategy of the computation, and we will present some of the results obtained
so far.

The organization of the paper is as follows: in section 2, we explain the
structure of contact terms in twisted ${\cal N}=2$ Yang-Mills theory. In section 3,
we show how the contact terms can be understood from the point of
view of Whitham hierarchies. In section 4, we study the connection
between the blowup formula in twisted Yang-Mills theory and the Toda-Whitham
approach.
In section 5, we review, following \cite{rf:our}, the uses of Whitham hierachies
in the problem of soft supersymmetry breaking in ${\cal N}=2$ theories. In section
6 we present some results on instanton expansions using the RG equations.
Finally, in section 7 we state some conclusions and open problems.

\section{Contact terms in twisted ${\cal N}$=2 Yang-Mills theory}
\setcounter{equation}{0}
\subsection{Twisted ${\cal N}=2$ Yang-Mills theory}
One of the most important aspects of ${\cal N}=2$ Yang-Mills theories is their
relation to Donaldson theory. It is a well-known fact that any
${\cal N}=2$ theory in four dimensions can be ``twisted" to obtain a topological
quantum field theory (TQFT), {\it i.e.} a quantum field theory whose correlation functions
are formally metric-independent (for a review of TQFT and the twisting procedure, see for example
\cite{rf:cordes,rf:lalo}.) In the case of ${\cal N}=2$ Yang-Mills, the correlation
functions of the twisted theory are in fact the famous Donaldson invariants of
four-manifolds \cite{rf:tqft}. In this section, we will consider some aspects of this twisted
theory for the gauge group $SU(N)$ on a simply-connected four-manifold $X$. In the following, $r=N-1$ will
denote the rank of the group.

The first thing to do is to identify the gauge-invariant
operators of the twisted theory. The twisted theory is characterized by the existence of a
BRST charge or topological charge $Q$ (which is in fact a particular combination of the
original supersymmetric charges of the ${\cal N}=2$ theory), and the operators of the twisted theory have
to be BRST invariant. These operators are called {\it observables}. The simplest observables
of the theory are precisely the Casimirs of the gauge
group, which we will take to be the elementary symmetric polynomials in
the eigenvalues of the complex scalar field $\phi$ of the ${\cal N}=2$ vector multiplet:
\be
{\cal O}_k={1\over k} {\rm Tr}\phi^k + {\rm lower \, order \, terms}, \,\,\ k=2, \dots, N.
\label{casimirs}
\ee
Starting with these operators, one can generate the rest of the observables through the
so-called {\it descent procedure}. To do this, one needs another operator in the
twisted theory, the descent operator $G_{\mu}$ which has spin one and comes from another
combination of the SUSY charges. Acting with $G_{\mu}$ $n$ times on the Casimir operators
${\cal O}_k$ one generates an $n$-form, The integration of this $n$-form on an $n$-cycle on $X$ is
another observable, called the $n$-th topological descendant of ${\cal O}_k$. As we are assuming that
the manifold $X$ is simply connected, the topological
descendants will be constructed with two-cycles $S$ in $X$:
\be
I_k(S)=\int_S G^2 {\cal O}_k = {1 \over k}\int_S {\rm Tr}(\phi^{k-1} F)+ \dots,
\label{descent}
\ee
where $F$ is the Yang-Mills field strength. The basic problem of the TQFT is to compute the
generating function for the correlators of observables:
\be
Z(p_k,f_k, S)=\langle \exp \bigl( \sum_k ( p_k {\cal O}_k + f_k I_k(S) )
\bigr) \rangle_{\rm twisted \, theory}.
\label{genfun}
\ee
Notice that, in general, $S$ will be an arbitrary linear combination of basic
two-cycles $S_i$, $i=1, \dots, b_2(X)$, {\it i. e.} $S=\sum_{i=1}^{b_2(X)} t_i S_i$, therefore
\be
I_k(S)=\sum_{i=1}^{b_2(X)} t_i I_k(S_i).
\label{sumtwo}
\ee
In total, we have $r \cdot b_2(X)$ independent operators $I_k(S_i)$.

The computation of (\ref{genfun}) can be done using the low energy exact
solution of ${\cal N}=2$, $SU(N)$ Yang-Mills theory. This solution is encoded
in the hyperelliptic curve \cite{rf:sun}
\be
y^2 = P^2(x,u_k) - 4\Lambda^{2N},
\label{hyper}
\ee
where  $P(x,u_k) = x^{N} - \sum_{k=2}^N u_kx^{N-k}$ is the
characteristic polynomial of $SU(N)$, and $u_k=\langle {\cal O}_k
\rangle$ are the VEVs of the
Casimir operators (\ref{casimirs}). This curve has genus $g=r$,
and, as explained in \cite{rf:toda,rf:todamar}, it can be
identified with the spectral curve of the $N$ site periodic Toda
lattice. Associated to this curve, there is a meromorphic
differential of the second kind, with a double pole at infinity,
that can be explicitly written as:
\be
dS_{SW} = P'(x, u_k) {x dx \over y}.
\label{swdiff}
\ee
This one-form satisfies the equation:
\be
{\partial dS_{SW} \over \partial u_{k+1} } = \omega_k,
\label{defds}
\ee
where $\omega_k=x^{k-1}dx/y$ is a basis of holomorphic differentials.
Let $\gamma^I$ and $\gamma_{D,I}$ denote a symplectic basis of homology cycles for this curve,
$I=1, \dots, r$.  The $a^I$ variables of the prepotential ${\cal F}(a^I, \Lambda)$,
for the duality
frame associated to the cycles $\gamma^I$,
are given by the integrals over these cycles of $dS_{SW}$
\be
a^I(u_k,\Lambda) ={1 \over 2\pi i }  \oint_{\gamma^I} {x P'(u_k,x)\ov
\sqrt{P^2(u_k) -
4 \Lambda^{2N}}
}dx.
\label{laa}
\ee
The same expression holds for the dual variables
$a_{D,I}$ with $\gamma_{D,I}$ instead of
$\gamma^I$. Finally, the effective gauge couplings $\tau_{IJ}$ are just the
components of the Riemann matrix associated to the hyperelliptic curve (\ref{hyper}).
A fundamental aspect of the low-energy description of ${\cal N}=2$ gauge theories
is that one has to choose a duality frame, which in the language of Riemann surfaces
can be understood as a choice of the symplectic basis $\gamma^I$, $\gamma_{D,I}$. Different duality
frames are related by symplectic transformations of the form
\be
\Gamma = \pmatrix{ A & B \cr
C & D} \in {\rm Sp}(2r, {\bf Z}).
\label{symple}
\ee

\subsection{The integral over the Coulomb branch}

The moduli space of the hyperelliptic curve (\ref{hyper}) is parametrized by
the VEVs of the Casimirs $u_k$. For some values of these VEVs, the curve will be singular.
The singular locus is precisely the divisor ${\cal D}$ defined by the vanishing locus of the
discriminant of (\ref{hyper}), $\Delta_{\Lambda}=0$. It is well known that on this divisor
there are extra BPS states that become massless. The Coulomb branch of the ${\cal N}=2$
theory is then defined as
\be
{\cal M}_{\rm Coulomb} = {\bf C}^r - {\cal D}.
\label{coulbra}
\ee
The basic principle to compute the generating function (\ref{genfun}) has been
introduced by Moore and Witten in \cite{rf:mw} and states
that this function is given by the sum of
two different contributions:
one comes from the Coulomb branch, and the other comes from the divisor $\cal D$:
\be
Z= Z_{\rm Coulomb} + Z_{\cal D}.
\label{basic}
\ee
The first piece is given by an integral over the Coulomb branch, while
the second piece involves, in general, Seiberg-Witten invariants associated to
the extra massless BPS states. In this paper, we will focus on $Z_{\rm Coulomb}$
following \cite{rf:moore1,rf:moore2}. As shown in \cite{rf:mw,rf:moore2}, this contribution
is different from zero only when $b_2^+(X)=1$. The explicit expression of this integral can be
worked out using the low-energy effective action encoded in (\ref{hyper}) and reads
\be
Z_{\rm Coulomb} = \int_{{\cal M}_{\rm Coulomb} }
[da  d\bar   a]
A(u_k )^\chi B(u_k)^\sigma  e^{\sum p_k u_k  + S^2\sum f_k f_l T_{k,l}} \Psi.
\label{coulint}
\ee
The integrand of (\ref{coulint}) has various ingredients. First of all, there is
a gravitational part first studied in \cite{rf:sdual} in the $SU(2)$ case, and generalized
in \cite{rf:moore1,rf:moore2,rf:lns} to simply-laced groups. This part involves the factors:
\beqa
A(u_k )^\chi &=& \alpha^{\chi}
\biggl( {\rm det} {\partial u_k \over \partial a^I} \biggr)^{\chi/2}, \nonumber\\
B(u_k)^\sigma &=& \beta^{\sigma} \Delta_{\Lambda}^{\sigma/8}.
\label{grav}
\eeqa
The first factor is a modular form of weight $(-\chi/2,0)$, and $B(u_k)$
is a modular invariant. In these equations, $\alpha$, $\beta$ are constants. Notice that
the quantities involved here are very natural quantities associated to the hyperelliptic
curve (\ref{hyper}), namely, the determinant of the matrix of periods of $\omega_k$, and
the discriminant of the curve.

Another ingredient in (\ref{coulint}) is $\Psi$, which is given by a sum over a
lattice $\Gamma$. In order to write this quantity, let's
introduce the following notation:
\be
V_I\equiv \sum_k f_k {\partial u_k \over \partial a^I}.
\label{twobs}
\ee
We also need some geometrical ingredients. As we are on a manifold of
$b_2^+(X)=1$, given a metric $g$ there exists a unique
anti-self-dual form $\omega \in H^{2,+}(X, {\bf R})$ such that $\omega^2=1$. The self-dual part of a
two-form is then given by $\lambda_+ =(\lambda,\omega)$, where $(\cdot, \cdot)$ is the usual
product in cohomology. The lattice
sum in $\Psi$ comes essentially from the evaluation of the partition function of the photons
in the effective $U(1)^r$ gauge theory. A topological sector in the effective theory is specified
by $r$ 2-forms $\lambda^I \in H^2 (X, {\bf R})$. The lattice $\Gamma$ consists of vectors of the form
\be
{\vec \lambda} = \sum_{I=1}^r \lambda^I \vec \alpha_I,
\label{lat}
\ee
where $\{\vec \alpha_I \}_{I=1, \dots, r}$ are the simple roots of $SU(N)$.\footnote{For simplicity, we
are assuming that there are no magnetic fluxes turned on. The general case is analyzed in
\cite{rf:moore2}.}
In terms of these quantities, $\Psi$ can be written as follows:
\beqa
\Psi& =&(\det {\rm Im} \tau)^{-1/2}
 \exp\bigl[   { 1 \over  8 \pi  }V_J [({\rm Im} \tau)^{-1}]^{JK} V_K  S_+^2 \bigr]\nonumber\\
&\times & \sum_{\vec \lambda \in \Gamma }
\exp\biggl[ - i \pi {\overline \tau}_{IJ} (\lambda_+^I,\lambda_+^J)
- i \pi   \tau_{IJ} (\lambda_-^I,\lambda_-^J)
- i \pi (\vec \lambda  \cdot  \vec \rho,  w_2 (X))  - i   V_I
(S,\lambda_-^I) \biggr] \nonumber\\
&\times &\int \prod_{I=1}^r d \eta^I d \chi^I
\exp\biggl\{ -{i  \sqrt {2} \over 16 \pi} {\overline {\cal F} }_{IJK}\eta^I
\chi^J [4\pi (\lambda_{+}^K, \omega)
+ i ({\rm Im} \tau)^ {KL} V_L (S, \omega) ]
 \nonumber\\
& &+ {1 \over 64 \pi } {\overline {\cal F}}_{KLI} ({\rm Im}  \tau)^{IJ}
{\overline {\cal F} }_{JPQ}
\eta^K  \chi^L \eta^P \chi^Q  \biggr\},
\label{latsum}
\eeqa
where ${\vec \rho}$ is the Weyl vector, $w_2(X)$ is the second Stiefel-Whitney
class of $X$, and $\eta^I$, $\chi^I$ are Grassmannian coordinates which arise from
the zero modes of the fermion fields in the theory. ${\cal F}_{IJK}$ denote the third
derivatives of the prepotential. Notice that $\Psi$ depends explicitly on
the metric of $X$ through the two-form $\omega$. It is precisely this dependence
which leads to the wall-crossing phenomena of Donaldson theory in manifolds
with $b_2^+(X)=1$, when the gauge group is $SU(2)$ \cite{rf:mw}.

In (\ref{coulint}) we have also included some terms of the form $T_{k,l}S^2$ in the exponential. These
terms, which are proportional to the intersection form of the two-cycles, are the {\it contact terms}
that we want to study. To understand the origin of these terms, notice that the operators $I_k(S)$ that
we have introduced in (\ref{sumtwo}) are non-local operators. These operators have a low-energy counterpart
${\widetilde I}_k(S)$ which can be obtained using the descent procedure in the effective
abelian theory, and have been included in $\Psi$.
But if we consider products of operators, and we go to the low-energy theory, we rather expect
an extra contribution localised on the intersection of the surfaces that support the operators:
\be
I_k(S_1) I_k(S_2) \rightarrow {\widetilde I}_k(S_1){\widetilde I}_k(S_2) + T_{k,l}(S_1 \cap S_2).
\label{lowcon}
\ee
This is precisely what we have taken into account in (\ref{coulint}) by introducing
the contact terms $T_{k,l}$.
It is important to notice that these terms are not predicted {\it a priori} by the
Seiberg-Witten solution for the low-energy effective action. One has to find extra
conditions in order to be able to find their structure.

\subsection{The structure of the contact terms}

To have more information on the contact terms, the first step is to write
the integrand of (\ref{coulint}) in a way that makes manifest the behavior under
symplectic transformations. One introduces, then, the following generalized
Siegel-Narain theta function:
\beqa
& & \Theta _{\Gamma}(\tau_{IJ}, \alpha_I, \beta^I; P, \xi_I)= \exp \biggl[ - i \pi
 (\alpha_I, \beta^I)
+ {\pi \over 2} \Bigl( \xi_{I,+} ({\rm Im} \tau)^{IJ} \xi _{J,+} - \xi_{I,-}
({\rm Im} \tau)^{IJ} \xi _{J,-} \Big)
\biggr] \nonumber\\
& & \,\,\,\,\,\,\,\,\,\ \times \sum_{\vec \lambda \in \Gamma}\exp \biggl[ - i \pi
{\overline \tau} _{IJ} ({\hat
\lambda}^I_+, {\hat \lambda}^J_+)
-  i \pi \tau_{IJ} ({\hat \lambda}^I_-, {\hat \lambda}^J_-) - 2 \pi i  ({\hat
\lambda}^I,  \xi_I)
+ 2\pi i ({\hat \lambda}^I, \alpha_I) \biggr] ,\nonumber\\
\label{thetarank}
\eeqa
where $ {\hat \lambda}^I = \lambda^I + \beta^I$.
If we take
\beqa
\xi_I &\equiv& {1 \over 2 \pi} V_I S_- + { {\sqrt 2} \over 16 \pi} {\overline{\cal
F}}_{IJK} \eta^J
\chi ^K   \omega, \nonumber\\
\beta^I &= &0, \,\,\,\,\,\,\ \alpha_I = {1 \over
2} w_2(X),
\,\,\,\ I = 1, \cdots ,r,
\label{varth}
\eeqa
the
lattice sum $\Psi$ can be written as
\beqa
\Psi &= & \exp \bigl[ {S^2 \over 8 \pi} V_I [({\rm Im} \tau)^{-1}]^{IJ}V_J \bigr]
(\det {\rm Im}
 \tau )^{-1/2}
\nonumber\\
& \times& \int  \prod_{I=1}^r d \eta^I d \chi^I \exp \bigl[ { {\sqrt 2} \over 16 \pi}
{\overline {\cal F}}_{IJK}
\eta^I \chi ^J ({\rm Im} \tau)^{KL} V_L (S, \omega) \bigr]
\Theta_{\Gamma}  (\tau_{IJ}, \alpha_I, \beta^I; P, \xi_I). \nonumber\\
\label{rewr}
\eeqa
It is easy to check \cite{rf:moore2} that $\widehat \Psi =
\exp \bigl[ -{S^2 \over 8 \pi}
V_I [({\rm Im} \tau)^{-1}]^{IJ}V_J \bigr]\Psi$
is a modular form
of weights $(b_2(X)/2, -1)$ (one has to take
into account that $b_2^+(X)=1$).
The Coulomb integral then reads:
\beqa
Z_{\rm Coulomb} = &\int_{{\cal M}_{\rm Coulomb}}[du d{\overline u}] \exp \biggl[ \bigl(
\sum f_k f_l T_{k,l} + {1 \over 8 \pi} V_I [({\rm Im} \tau)^{-1}]^{IJ}V_J
\bigr) S^2 \biggr]\nonumber\\
&\,\,\,\,\,\,\,\ \times \bigl|{\rm det} \bigl({\partial a^I \over
\partial u_k}\bigr) \bigr|^2 A(u_k)^{\chi} B(u_k) ^{\sigma} \widehat \Psi.
\label{final}
\eeqa
The factor in the second line of (\ref{final}) is a modular invariant (notice that, as $X$
is simply connected, $\chi=2+b_2(X)$). We then see that
this expression for the generating function
(except for the exponential involving $S^2$)
is the integral of a duality invariant object over a moduli space parametrized
by the VEVs of the Casimirs, which are duality invariant coordinates.
This is precisely what
we expect on physical grounds: the generating function is a physical quantity
and cannot depend on the choice of duality frame in the effective theory.
This argument in fact forces the exponent in $S^2$ to be a modular invariant
as well, and this gives the first constraint on the contact terms: the quantity
\be
T_{k,l} + {1 \over 8 \pi} {\partial u_k \over \partial a^I}[({\rm Im} \tau)^{-1}]^{IJ}
{\partial u_l \over \partial a^J}
\label{firstcon}
\ee
must be invariant under the action of the symplectic group ${\rm Sp}(2r,{\bf Z})$.
The other constraint on the contact terms has to do with their physical origin: these
terms are quantum corrections and vanish at tree level, therefore they have to
go to zero in the semiclassical
region of moduli space ({\it i.e.} when $\Lambda/a^I \rightarrow 0$). The duality
transformation of the second term in (\ref{firstcon}) is easily worked out, and one finds
that under a duality transformation it is shifted by:
\be
- {i\over 4\pi} {\partial u_k \over \partial a^I}
\bigl[ (C\tau + D)^{-1} C\bigr]^{IJ}{\partial u_l \over \partial a^J}.
\label{shifteq}
\ee
The transformation of the contact term
should compensate for this shift. Summarizing the discussion so far, we have found the
following constraints for the contact terms:

$\bullet$ They transform under an element of ${\rm Sp}(2r, {\bf Z})$ as follows:
\be
T_{k,l} \rightarrow T_{k,l} + { i \over 4 \pi}{\partial u_k \over \partial a^I}
\bigl[ (C\tau + D)^{-1} C\bigr]^{IJ}{\partial u_l \over \partial a^J}.
\label{contrans}
\ee

$\bullet$ $T_{k,l} \rightarrow 0$ semiclassically.

As shown in \cite{rf:mw}, these two properties determine the contact terms unambiguously.
The problem, of course, is to find explicit expressions for them.

In the case of the pure Yang-Mills theory with gauge group $SU(2)$, the only contact term is $T_{2,2}$.
The solution to the constraints
above was found in \cite{rf:mw}, and the result was expressed in terms of the Eisenstein series $E_2(\tau)$.
A general procedure to find $T_{2,2}$ was described in \cite{rf:moore1,rf:moore2} by using the so-called RG equation of
${\cal N}=2$ theories \cite{rf:matone,rf:cobi}, as well as some aspects of the formalism developed in
\cite{rf:soft,rf:moresoft,rf:luisIyII}. The RG equation for ${\cal N}=2$ theories states that
\cite{rf:matone,rf:cobi}
\be
{\partial {\cal F} \over \partial \tau_0} = {1 \over 4} u_2,
\label{rgeq}
\ee
where $\tau_0$ is defined as follows: for asymptotically free theories,
$\Lambda^{\beta}= {\rm e}^{\pi i \tau_0}$, where $\beta$ is the coefficient of the
one-loop beta function (for example, $\beta=2N$ for $SU(N)$). For the self-dual theories,
$\tau_0$ is the microscopic coupling constant. The key point is to consider now the
second derivatives of the prepotential. On the one hand, from (\ref{rgeq}) one has that
\be
{\partial^2 {\cal F} \over \partial a^I \partial \tau_0}= {1 \over 4}{\partial
u_2 \over \partial a^I}.
\label{firstder}
\ee
On the other hand, the fact that $\tau_0$ is invariant under duality transformations
implies the following transformation law \cite{rf:moresoft}:
\be
{\partial ^2 \F \over \partial \tau_0^2} \rightarrow
{\partial ^2 \F \over \partial \tau_0^2}-{\partial^2 \F \over \partial \tau_0
\partial a^I}
[(C \tau + D)^{-1}]^I_{~J} C^{JK} {\partial^2 \F \over \partial \tau_0
\partial a^K}.
\label{dualct}
\ee
This is precisely the shift we found in (\ref{contrans}) under symplectic transformations.
It is easy to see that ${\partial ^2 \F / \partial \tau_0^2}$ is zero semiclassically
(this is related to the fact that the classical prepotential is linear in the gauge
coupling constant). It follows that the contact term associated to the quadratic
Casimir can be written as
\be
T_{2,2}  = { 4 \over \pi i}  {\partial ^2 \F \over  \partial \tau_0^2}.
\label{contact}
\ee
$T_{2,2}$ can be in fact evaluated very explicitly in the $SU(2)$ case \cite{rf:moore1,rf:luisIyII}. For
the theories with $N_f\le 3$, one finds:
\be
T_{2,2} = -{1 \over 24} E_2 (\tau) \bigl( {du \
\over da} \bigr)^2 + {1\over 3} \bigl( u + \delta_{N_f,3}{\Lambda_3^2 \over 64} \bigr),
\label{contnf}
\ee
and for the $N_f=4$ case one has:
\be
T_{2,2} = -{1 \over 24} E_2 (\tau) \bigl( {du \
\over da} \bigr)^2 + E_2 (\tau_0) {u\over 3} + {1 \over 9} R E_4 (\tau_0),
\label{conself}
\ee
where $R= \sum_{a=1}^4 m_a^2/2$, and $E_2$, $E_4$ are the normalized Eisenstein series.
This provides an elegant
solution to our problem for this particular contact term. The authors of \cite{rf:lns}
found equivalent expressions for the $SU(2)$ theories in the massless case
and for $N_f \le 3$.

The approach based on the RG equation suggests how to find $T_{k,l}$ for any $k,l$.
If one is able to find some additional variables $T_n$, $n=2, \dots, N-1$ in the prepotential,
which are invariant under duality and such that
\be
{\partial \F \over \partial T_n} \sim u_{n+1},
\label{simil}
\ee
then ${\partial ^2 \F / \partial T_k\partial T_l}$ will have, essentially, the
duality properties of the contact term $T_{k+1, l+1}$, and one is halfway to the solution
of the problem. The two obvious questions are: Can we find an explicit construction
of the generalized variables $T_n$?, and: Can we compute the second derivatives
${\partial ^2 \F / \partial T_k\partial T_l}$ in terms of elementary data associated
to the Seiberg-Witten solution? These two questions are answered in \cite{rf:ITEP}
in the affirmative through the use of the Whitham hierarchy approach to the
prepotential, which will be the subject of the next section.

\section{Whitham hierarchies and contact terms}

\subsection{Whitham hierarchy and Seiberg-Witten solution}

The approach to the theory of the prepotential based on Whitham hierarchies has been
developed in \cite{rf:krich,rf:prepw,rf:im}. In \cite{rf:ITEP} it was shown that, using this approach, one
can derive RG equations with the structure (\ref{simil}). We will follow here \cite{rf:ITEP}
and also \cite{rf:our}.

The usual Seiberg-Witten differential is a meromorphic differential with a second-order
pole at infinity, and such that its variations with respect to the moduli $u_k$ are holomorphic
differentials. In order to deform the Seiberg-Witten theory and to embed it in a
larger framework, one considers a series of meromorphic
differentials $d\widehat \Omega_n$ (in the notation of \cite{rf:ITEP}),
with poles of order $n+1$ at infinity and satisfying the condition
\be
{\partial d\widehat \Omega_n \over \partial {\rm moduli}} = {\rm holomorphic}.
\label{holo}
\ee
One then introduces a generating functional for these one-forms with auxiliary
parameters $T_n$, $n\ge 1$:
\be
dS =\sum_{n\ge 1} T_n d\widehat \Omega_n.
\label{gentimes}
\ee
The parameters $T_n$ are precisely the slow times of the Whitham hierarchy. One of the
results of \cite{rf:ITEP} is an explicit expression for the meromorphic one-forms
$d\widehat \Omega_n$:
\be
d\widehat \Omega_n =P^{n/N}_+(x) {P'(x) dx \over {\sqrt {P^2-4}}},
\label{expliom}
\ee
where $\left(\sum_{k=-\infty}^\infty c_k\lambda^k\right)_+ =
\sum_{k=0}^\infty c_k\lambda^k$. It is easy to check that
$P_+^{1/N}(x)=x$, therefore $dS (T_1, T_{n\ge 2}=0)= T_1 dS_{SW}(\Lambda=1)$. We then see
that we recover the usual Seiberg-Witten differential when $T_{n \ge 2}=0$. Starting with the
enlarged differential (\ref{gentimes}), one can construct a deformation of the usual Seiberg-Witten
theory. For instance, one defines the periods of $dS$ as
\beqa
\alpha^I(u_k,T_1,T_2,...) &=&  \sum_{n\geq 1} {T_n \over 2\pi i }\oint_{\gamma^I}
d\widehat\Omega_n
\cr
&=&
\sum_{n\geq 1} {T_n \over 2\pi i } \oint_{\gamma^I}{P(u_k)^{n/N}_+ P'(u_k)\ov
\sqrt{P^2(u_k) - 4} }
    d\lambda \cr
&=& T_1 a^I(u_k,\Lambda=1) + {\cal O}(T_{n>1}),
\label{alpha}
\eeqa
where $a^I(u_k,\Lambda)$ is the usual Seiberg-Witten period of $dS_{SW}$.
In the same way, one defines the $\alpha_{D\,I}$ with the expression (\ref{alpha})
but with the $\gamma_{D,I}$ replacing the $\gamma^I$. Following now the usual steps
in rigid special geometry, we obtain an enlarged prepotential ${\cal F} (\alpha^I, T_n)$,
taking $\alpha^I$ and the slow times as independent variables.

We will now make more
precise the relation with the usual Seiberg-Witten solution, following
\cite{rf:our}. We define the rescaled times $\hat T_n$ and VEVs $\hat u_k$ as folows:
\be
\hat T_n = T_n/T_1^n, \,\,\,\,\,\,\  \hat u_k = T_1^k u_k,
\label{cambio}
\ee
with $\hat T_1 = 1$. It is easy to see that (\ref{alpha}) can be written as
\be
\alpha^I(u_k,T_1,\hat T_2,\hat T_3,...) = \sum_{n\geq 1} {\hat T_n \over 2\pi i
}
\oint_{\gamma^I}{P^{n/N}_+(\hat u_k) P'(\hat u_k)\ov \sqrt{P^2(\hat u_k) - 4T_1^{2N}}
}
    d\lambda.
\label{hatalpha}
\ee
In particular, after setting $\hat T_2 = \hat T_3 =...= 0$
we find that
$$
\alpha^I(u_k,T_1,\hat T_{n>1} = 0) = T_1 a^I(u_k,1) =
a^I(\hat u_k,\Lambda=T_1) ~.
\label{alaa}
$$
In conclusion, we may identify $T_1$ with $\Lambda$ in the submanifold
$\hat T_2 =
\hat T_3 = ...= 0$, provided the moduli space is parametrized with $\hat u_k$
instead of  $u_k$. This shows that the Whitham hierarchy approach to the prepotential
gives a deformation of the usual Seiberg-Witten theory, in such a way that
the quantum scale can be identified with the first slow time of the
Witham hierarchy.

\subsection{Derivatives of the prepotential and contact terms}

As we noticed in the last section, the slow times would be useful
for the problem of the contact terms if they were ``dual" to the higher Casimirs,
in the sense of (\ref{simil}). We should then compute the
derivatives of the prepotential $\F(\alpha,T)$. This computation
was the main result of \cite{rf:ITEP}. One of the basic ingredients
in the answer is the Riemann theta function
$\Theta[\vec \alpha, \vec \beta](\vec \xi|\tau)$ and characteristics
$\vec \alpha=(\alpha_1, \dots, \alpha_r)$,
$\vec
\beta=(\beta_1, \dots, \beta_r)$:
\be
\Theta [\vec \alpha, \vec \beta] (\vec \xi|\tau)= \sum_{n_K \in {\bf Z} } \exp
\bigl[ i \pi \tau_{IJ}(n_I+ \beta_I)
(n_J+ \beta_J) + 2\pi i (n_I+ \beta_I)(\xi_I + \alpha_I) \bigr].
\label{thetaf}
\ee
The derivatives of the prepotential turn out to be:
\beqa
\pder{\F}{T_n} &=& {\beta\ov 2\pi i n} \sum_{m} m T_m\H_{m+1,n+1} ~, \nonumber
\\
\ppder{\F}{\alpha^I}{T^n} &=& {\beta\ov 2\pi i n} \pder{\H_{n+1}}{a^I} ~,
\label{completeness} \\
\ppder{\F}{T_m}{T_n} &=& -{\beta\ov 2\pi i}\left(
\H_{m+1,n+1} + {\beta\ov mn}\pder{\H_{m+1}}{a^I}\pder{\H_{n+1}}{a^J}
{1\ov i\pi}\d_{\tau_{IJ}}\log \Theta_E(0|\tau)\right) ~.\nonumber
\eeqa
In these equations, $\Theta_E(0|\tau)$ denotes Riemann's theta function
with a certain characteristic $E$, evaluated at the origin;
$\beta= 2N,~m,n = 1,...,r=N-1$, and
derivatives with respect to $T_n$ are taken at constant $\alpha^I$.
According to \cite{rf:ITEP},
the characteristic $E$ appearing in
(\ref{lasecu})--(\ref{lasecudef}) is an even, half-integer characteristic
associated to a particular splitting of the roots of the discriminant.
Notice that the characteristic $E$ depends on the duality frame. An explicit
expression for $E$ in the electric frame will be given in the next section.
The functions $\H_{m,n}$ are certain homogeneous combinations of the
Casimirs $u_k$, given by
$$
\H_{m+1,n+1} = {N\ov mn} \hbox{res}_\infty \left(
P^{m/N}(x)dP_+^{n/N}(x) \right) = \H_{n+1,m+1} ~,
$$
and $$\H_{m+1} \equiv \H_{m+1,2} = {N\ov m} \hbox{res}_\infty P^{m/N}(x)
dx = u_{m+1} + {\cal O}(u_m).$$
Here res$_P$ stands for the usual Cauchy residue at the point $P$.
We have for instance $\H_{2,2}
= \H_{2} = u_2,~\H_{3,2} =\H_{3}= u_3$ and $\H_{3,3} = u_4+ {N-2\over 2N}
u_2^2$.
As we have seen, in order to recover the Seiberg-Witten solution, it is better
to use the rescaled variables $(\Lambda, \hat T_{n\ge 2})$. It is straightforward to
reexpress
all the formulae in (\ref{completeness}) as
derivatives of $\F$ with respect to $\alpha^I, \hat T_n$
and $\Lambda$. Most of the factors $T_1$ can be used to promote
$u_k$ to $\hat u_k$ or, rather, to the homogeneous combinations thereof:
\be
\hat \H_{m+1,n+1} = T_1^{m+n}\H_{m+1,n+1} ~~~~ \Rightarrow ~~~~
\hat \H_{m+1} = T_1^{m+1}\H_{m+1}
\ee
(since $\H_{m+1}= \H_{m+1,2}$).
The remaining $T_1$'s are absorbed in making up
$\hat a^I
\equiv T_1a^I(u_k,1) = a^I(\hat u_k,T_1)$.
Altogether we find
\beqa
\pder{\F}{\log\Lambda\,} & = & {\beta\ov 2\pi i} \sum_{n,m\geq 1}
m\hat T_m\hat T_n\hat\H_{m+1,n+1} ~~, ~~~~~~~~~ \pder{\F}{\hat T_n}
{}~=~ {\beta\ov 2\pi i n} \sum_{m\geq 1} m\hat T_m \hat\H_{m+1,n+1} ~,
\nonumber \\
\ppder{\F}{\alpha^I}{\log \Lambda} & = & {\beta\ov 2\pi i} \sum_{m\geq
1}\hat T_m\pder{\hat\H_{m+1}}{\hat a^I} ~~, ~~~~~~~~~~~~~~~
\ppder{\F}{\alpha^I}{\hat T_n} ~=~ {\beta\ov 2\pi i
n}\pder{\hat\H_{n+1}}{\hat a^I} ~, \nonumber \\
\dpder{\F}{(\log\Lambda)} & = & - {\beta^2\ov 2\pi i}\sum_{m,n\geq 1}
\hat T_m  \hat T_n \pder{\hat\H_{m+1}}{\hat a^I} \pder{\hat\H_{n+1}}{\hat
a^J}\, {1\ov i\pi} \d_{\tau_{IJ}}\log\Theta_E(0|\tau) ~, \label{lasecu} \\
\ppder{\F}{\log\Lambda\,}{\hat T_n} & = & - {\beta^2\ov 2\pi i
n}\pder{\hat\H_{n+1}}{\hat a^I} \sum_{m\geq 1}\hat T_m
\pder{\hat\H_{m+1}}{\hat a^J}\, {1\ov i\pi}
\d_{\tau_{ij}}\log\Theta_E(0|\tau) ~, \nonumber \\
\ppder{\F}{\hat T_m}{\hat T_n} & = & - {\beta\ov 2\pi i} \left(
\hat\H_{m+1,n+1}+{\beta\ov mn} \pder{\hat\H_{m+1}}{\hat
a^I}\pder{\hat\H_{n+1}}{\hat a^J}{1\ov i\pi}
\d_{\tau_{IJ}}\log\Theta_E(0|\tau) \right) \nonumber ~,
\eeqa
with $m,n\geq 2$. In these expressions $\hat T_1 = 1$.
The restriction to the submanifold
$\hat T_2 = \hat T_3=...=0$ yields formulae which are expressed
in terms of the original Seiberg-Witten data. Notice that in this subspace
$\alpha^I(u_k,T_1,\hat T_{n>1} = 0)=
\hat a^I$, hence
\beqa
\pder{\F}{\log\Lambda\,} & = & {\beta\ov 2\pi i } \hat\H_{2} ~~,~~~~~~~~
{}~~~~~~~~~~~~~~~~~~ \pder{\F}{\hat T_n}
{}~=~ {\beta\ov 2\pi i n}  \hat\H_{n+1} ~, \nonumber \\
\ppder{\F}{\hat a^I}{\log\Lambda} & = & {\beta\ov 2\pi i}\pder{\hat\H_{2}}{\hat
a^I} ~~,~~~~~~~~~~~~~~~~~~~ \ppder{\F}{\hat a^I}{\hat T_n} ~=~ {\beta\ov 2\pi i
n}\pder{\hat\H_{n+1}}{\hat a^I} ~, \nonumber \\
\dpder{\F}{(\log\Lambda)} & = & -{\beta^2\ov 2\pi i}
\pder{\hat\H_{2}}{\hat a^I}\pder{\hat\H_{2}}{\hat a^J}{1\ov i\pi}
\d_{\tau_{IJ}}\log\Theta_E(0|\tau) ~, \label{lasecudef} \\
\ppder{\F}{\log\Lambda\,}{\hat T_n} &=& -{\beta^2\ov 2\pi i n}
\pder{\hat\H_{2}}{\hat a^I}\pder{\hat\H_{n+1}}{\hat a^J}{1\ov i\pi}
\d_{\tau_{ij}}\log\Theta_E(0|\tau) ~, \nonumber \\
\ppder{\F}{\hat T_m}{\hat T_n} &=& -{\beta\ov 2\pi i} \left(
\hat\H_{m+1,n+1}+{\beta\ov mn}
\pder{\hat\H_{m+1}}{\hat a^I}\pder{\hat\H_{n+1}}{\hat a^J}{1\ov i\pi}
\d_{\tau_{IJ}}\log\Theta_E(0|\tau) \right) ~, \nonumber
\eeqa
with $m,n\geq 2$. We will always work with the coordinates $\Lambda$ and $\hat T_n$
defined in (\ref{cambio})
and, therefore, hats will be omitted everywhere.

The first derivatives of the prepotential with respect to the slow
times can be interpreted as generalized RG equations.
These equations show that, indeed, the slow times $T_n$ are dual to the
Casimirs ${\cal H}_{n+1}$, at least in the submanifold $T_{n\ge 2}=0$.
One can also work out the transformation properties under the symplectic
group of the second derivatives of the prepotential. To do this, notice that
the quantities ${\cal H}_{n,m}$ are combinations of gauge invariant VEVs,
therefore they are invariant under duality. The slow times are deformation
parameters of the theory, and they are also invariant. The symplectic transformation
of the periods $\alpha^I$, $\alpha_{D,I}$ are the usual ones for the
variables $a^I$, $a_{D,I}$ of the Seiberg-Witten theory, and one derives the
transformation law for the prepotential obtained in
\cite{rf:matone,rf:cobi}
\be
\F^{\Gamma} =
\F + {1\ov 2} \alpha^I(D^T B)_{ij} \alpha^J +
  {1\ov 2} \alpha_{D\,I}(C^T A)^{IJ} \alpha_{D,J} +
      \alpha^I (B^T C)_I{^J} \alpha_{D,J}.
\label{gprepot}
\ee
One then finds \cite{rf:our},
\be
{\partial ^2 \F \over \partial T_n \partial T_m} \rightarrow
{\partial ^2 \F \over \partial T_n \partial T_m}-{\partial^2 \F \over \partial T_n
\partial \alpha^I}
[(C \tau + D)^{-1}]^I_{~J} C^{JK} {\partial^2 \F \over \partial T_m
\partial \alpha^K}.
\label{dualctgen}
\ee
This transformation law can also be checked on the explicit expressions in
(\ref{lasecu}), by using the transformation
law of the theta function under a symplectic transformation:
\be
\Theta [\vec \alpha, \vec \beta] (\vec \xi|
\tau)\rightarrow
{\rm e}^{i \phi} ({\rm det} (C\tau + D))^{1/2}
\exp \bigl[ \pi i \xi^t (C\tau + D)^{-1} C \xi \bigr] \Theta [
\vec \alpha ,\vec  \beta](\vec \xi|\tau),
\label{thetatrans}
\ee
where ${\rm e}^{i \phi}$ is an eighth root of unity. The transformation law
(\ref{dualctgen}) is the generalization of (\ref{dualct})
that we were looking for. We would be
tempted to identify the contact term $T_{k+1,l+1}$ with the second derivative
${\partial ^2 \F / \partial T_{k} T_{l}}$, up to a normalization factor.
There are, however, two subtleties one
has to take into account in order to make this identification. The first one
is that, in the above model, the Casimirs
are not $u_k$, but the homogeneous combinations ${\cal H}_k$. Therefore, to identify the
contact terms with the second derivatives of the prepotential, we have to define the
operators $I_k (S)$ starting with ${\cal H}_k$
instead of the $u_k$ in the descent procedure. This is simply a redefinition of
which gauge-invariant operators
we take as a basis to construct the observables. The second subtlety is that
the contact terms are not only determined by their transformation properties.
As we discussed in the previous section, there is the extra constraint that
$T_{k,l}$ vanishes semiclassically. Let's then analyze
the semiclassical behavior of ${\partial ^2 \F / \partial T_{k} T_{l}}$. If we use the
semiclassical expression for the effective gauge coupling $\tau_{IJ}$, we find that:
\be
\d_{\tau_{IJ}}\log\Theta_E(0|\tau) \sim {\cal O} \biggl(
{\Lambda^{2N} \over Z_{\vec \alpha_+} ^{2N}} \biggr),
\label{thetavan}
\ee
where $Z_{\vec \alpha_+}=\vec a \cdot \vec \alpha_+$, $\vec a= \sum_I  a^I \vec\alpha_I$,
and $\vec \alpha_+$ are the positive roots. Therefore, the term in
${\partial ^2 \F / \partial T_{k} T_{l}}$ involving this derivative will vanish in the
semiclassical region $(\Lambda / Z_{\vec \alpha_+})\rightarrow 0$, for $k,l=1, \dots, N-1$
(we take $T_1 = \Lambda$ here). On the other hand, ${\cal H}_{n+1,m+1}$ does
not vanish. Therefore, ${\partial ^2 \F / \partial T_{k} T_{l}}$ does not have
the required behavior for $k,l=2, \dots, N-1$. This can be cured by a redefinition of the
prepotential. Consider the ``reduced" prepotential
\be
\F^{\rm red}= \F + {\beta \over 4 \pi i
} \sum_{n,m\ge2}
T_n T_m \H_{n+1, m+1}.
\label{reduced}
\ee
The second derivatives of this prepotential with respect to the slow times are,
in the Seiberg-Witten submanifold $T_{n\ge 2}=0$,
\be
\bigl( {\partial^2 \F^{\rm red} \over
\partial T_{n} \partial T_{m}}\bigr)_{T_{n\ge 2}=0}=-{\beta^2\ov 2\pi i n}
\pder{\hat\H_{n+1}}{\hat a^I}\pder{\hat\H_{m+1}}{\hat a^J}{1\ov i\pi}
\d_{\tau_{IJ}}\log\Theta_E(0|\tau),
\label{reder}
\ee
and the contact terms are given by
\be
T_{k+1,l+1} = { \pi i kl \over \beta^2}\bigl( {\partial^2
\F^{\rm red} \over \partial T_{k} \partial T_{l}} \bigr)_{T_{n \ge 2} =0}.
\label{finct}
\ee
Notice that the piece that we substracted in the prepotential has a nonzero
classical limit, while (\ref{reder}) is a quantum correction. In other words, the
classical limit of the original prepotential $\F$ includes deformations which are
quadratic in the slow times. This is not what we expect in the problem of
evaluating the generating function (\ref{genfun}).
In this generating function, all the deformations are linear in the sources, and the quadratic terms
in these sources that appear in the effective theory (the contact terms) are only due to quantum effects.

To summarize, we have seen that the prepotential defined in terms of the
Whitham hierarchy is the appropriate object to understand the contact terms
of the twisted theory. However, the above remarks suggest that there should be
a way to improve the theory of the prepotential, in such a way that one obtains
the reduced version introduced in (\ref{reduced}), and that the slow times $T_n$
can be associated to any homogeneous combination of the Casimirs of degree $n+1$.

\section{Blowup formula}
In this section, we will discuss a further connection between
integrable systems and
the Coulomb integral, which appears when one considers the blowup formula. There are three
reasons why such a formula is interesting: first of all,
as shown in \cite{rf:lns}, the behavior under
blowup can be used to derive the contact terms. Second, as
shown in \cite{rf:moore2} and further
discussed in \cite{rf:taka}, the blowup formula involves in a direct way the tau function of the periodic Toda
lattice which underlies the Seiberg-Witten curve. Finally,
using the blowup formula one can
fix the characteristic $E$ which appears in (\ref{completeness}) \cite{rf:our}.
Of course, this formula is also very important in the context of
Donaldson theory, as it makes possible
to relate the Donaldson invariants of a manifold and its blowup
\footnote{It is interesting to
notice that the derivation of this formula in \cite{rf:fs} was
one of the first hints of a relation
between Donaldson theory and elliptic curves.}. This formula
was obtained using
the $u$-plane integral in \cite{rf:mw}, for the case of
$SU(2)$, and generalized to the higher rank case
in \cite{rf:moore2}.

Suppose that we have a four-manifold $X$, and
we consider the blownup manifold at a point $p$, $\hat X = {\rm Bl}_p(X)$.
Under this operation, the
homology changes as follows \cite{rf:gh}:
\be
H_2(X) \rightarrow H_2 (\hat X)=H_2 (X) \oplus {\bf Z} \cdot B,
\label{homol}
\ee
where $B$, the class of the exceptional divisor, satisfies
$B^2=-1$. As $b_2^+(\hat X) = b^+_2(X)$,
it makes sense to ask how the intergrand of $Z_{\rm Coulomb}$
changes under the blowup. First of all,
the blownup manifold $\hat X$ has an extra two-homology class,
therefore there are extra operators $I_k(B)$
that must be included in the generating function. We will
then write $\hat S= S+ t B$.
We want to compute then:
\be
\biggl\langle \exp\bigl[ \sum_k ( f_k I_k (S) +  t_k I_k(B) +
p_k {\cal O}_k) \bigr] \biggr\rangle_{\tilde{X}
},
\label{genblow}
\ee
in terms of correlation functions of the twisted theory on $X$. In this equation, $t_k \equiv f_k t$.
The first thing to do is to analyze the change
in the integrand under blowup. This is easy to obtain if we also assume that the metric is such
that $(B, \omega)=0$. In this case,
the lattice sum changes as follows:
\be
\Psi_{\hat X}=\biggl( \sum_{n^I} \exp \bigl[ \pi i \tau_{IJ}n^I n^J + i \sum_k t_k {\partial u_k \over \partial a^I}
n^I -i\pi \sum_I n^I \bigr]
\biggr) \Psi_{X},
\label{psich}
\ee
and we see that
\be
\Psi_{\hat X} = \Theta [\vec \Delta, 0] (\vec \xi |\tau) \Psi_X,
\label{psith}
\ee
where $\Theta [\vec \Delta, 0] (\vec \xi |\tau)$ is a theta function with
\be
\xi_I = \sum_k {t _k \over 2 \pi} {\partial u_k \over \partial a^I}, \,\,\,\,\,
\vec \Delta=(1/2, \dots, 1/2).
\label{vecth}
\ee
Notice that we have extracted from the
Siegel-Narain theta function a standard theta function on the hyperelliptic
curve (\ref{hyper}). The above expression is
valid in the electric frame, and the characteristic of the theta function
is inherited from the term $(\vec \rho \cdot \vec \lambda, w_2(X))$ in (\ref{latsum}).

Let's now analyze the measure in the integrand.
As $\chi(\hat X) = \chi (X)+1$ and $\sigma (\hat X) =
\sigma (X)-1$, the measure picks an extra factor
\be
\biggl( \det {  \partial u_J \over  \partial a^I} \biggr)^{1/2} \Delta_\Lambda^{-1/8} =
{ 1 \over \Theta[\vec \Delta, 0] (0|\tau)},
\label{factor}
\ee
as a consequence of Thomae formulae. Putting all these factors together, we see
that the blowup factor in the integrand is given (up to a constant) by:
\be
\tau (t_k|u_k)={\rm e}^{-\sum t_k t_l T_{k,l}} {\Theta[\vec \Delta,0](\vec \xi|\tau)\over
\Theta [\vec \Delta,0](0|\tau)}.
\label{blfactor}
\ee
As (\ref{blfactor}) is an extra factor in the integrand of $Z_{\rm Coulomb}$, it follows by our
arguments above that it must be invariant under duality transformations.
Using (\ref{thetatrans}), one can easily prove that
the duality invariance of the blowup factor fixes
the duality transformation of the contact terms,
and one finds (\ref{contrans}) again. The reason for the notation in (\ref{blfactor}) is that the
blowup factor is essentially the tau function of the Toda hierarchy (see, for example,
\cite{rf:ut}). Notice that, in this identification, the coupling constants
$t_k$ are interpreted as the {\it fast} times
of the hierarchy. Recall that there are two points of contact between integrable systems and the
Seiberg-Witten solution: one is the fact that the hyperelliptic curves are spectral curves of periodic
Toda chains (for pure Yang-Mills), and the other is that the prepotential is the logarithm of the
{\it quasiclassical} tau function. Now, we see that the tau function itself appears in the
twisted version of the theory as a blowup factor. This also gives a very direct link
between Donaldson theory and its generalizations,
and the hyperelliptic curve encoding the Seiberg-Witten
solution.

In principle, the blowup factor depends on
the moduli of the hyperelliptic curve in a complicated
way. However, it turns out that
the blowup factor only depends
on the fast times $t_k$ and on the Casimirs, as
we have indicated with our notation. More precisely,
let $\vec n =(n_2,\dots, n_N)$
be a vector of nonnegative integers, with
$|\vec n| = n_2 + \dots + n_N$. Then, there are {\it polynomials}
${\cal B}_{\vec n} (u_2, \cdots, u_N)$ in the
Casimirs such that:
\be
\tau (t_k|u_k)=\sum_{|\vec n|\ge 0} t_2^{n_2}
\cdots t_N^{n_N}{\cal B}_{\vec n} (u_2, \dots, u_N).
\label{expansion}
\ee
The physical reason for this, as explained in \cite{rf:mw}, is that one can interpret the blowup
as a local ``defect" created by an analog of the puncture operator. The effect of this puncture
associated to the exceptional divisor should be represented by an infinite number of local observables.
The ring of local observables is in fact generated by $u_2,\dots, u_N$, and an identity like (\ref{expansion})
should be valid. One can indeed prove (\ref{expansion}) by using duality invariance,
$R$-charge arguments and regularity \cite{rf:moore2}.
A consequence of (\ref{expansion}) is the following relation between the generating functions:
\be
\biggl\langle \exp\bigl[ \sum_k ( f_k I_k(S) + t_k I_k(B) + p_k {\cal O}_k) \bigr]
 \biggr\rangle_{\tilde{X}} =   \biggl\langle
\exp\bigl[ \sum_k (f_k I_k (S) +
  p_k {\cal O}_k )\bigr]   \tau(t_k|{\cal O}_k)
   \biggr\rangle_{X}\nonumber.
\label{relatione}
\ee
Notice that the blowup factor $\tau(t_k|u_k)$ does not depend on the manifold $X$. We have
then proved that the blowup changes the generating function by a
universal factor which can be expressed in terms of the zero observables.

As a corollary of (\ref{expansion}), we can derive an explicit expression
for the contact terms. The fact that the behavior under blowup
can be used to obtain the contact terms was first remarked in \cite{rf:lns}.
To obtain the expression, we simply expand (\ref{blfactor})
to second order in $t_k$. The first derivative of the theta function
is zero due to the choice of characteristic. We then find:
\be
\tau (t_k|u_k) = 1  - \sum_{k,l} \biggl( T_{k,l} + {1 \over 2 \pi i } \partial_{\tau_{IJ}}
\log \Theta[\vec \Delta,\vec 0] (0|\tau)
{\partial u_k \over \partial a^I}{\partial u_l \over \partial a^J} \biggr)
 t_k t_l + \cdots,
\label{blowup}
\ee
Because of (\ref{expansion}), this means that
\be
T_{k,l}=-{1 \over 2 \pi i }\partial_{\tau_{IJ}}
\log \Theta[\vec \Delta,\vec 0] (0|\tau){\partial u_k \over \partial a^I}{\partial u_l \over \partial a^J}
+ {\cal B}_{\vec n_{k,l}}(u_2,
\dots, u_N),
\label{conblow}
\ee
where $\vec n_{k,l}$ are the vectors with $|\vec n_{k,l}|=2$,
corresponding to the quadratic terms in (\ref{expansion}).
The requirement that $T_{k,l}$ vanishes semiclassically implies that ${\cal B}_{\vec n_{k,l}}(u_2,
\dots, u_N)=0$, and we finally find:
\be
T_{k,l}=-{1 \over 2 \pi i }\partial_{\tau_{IJ}}
\log \Theta[\vec \Delta,\vec 0] (0|\tau){\partial u_k \over \partial a^I}{\partial u_l \over \partial a^J},
\label{confin}
\ee
which is the expression found in \cite{rf:lns}. If we compare now to (\ref{finct}), we
see that they have the same structure. The only difference is due to the fact that the expression in (\ref{finct})
is valid if the descent operators are constructed with the operators ${\cal H}_k$. Here, instead, we
are considering the operators that we defined in (\ref{descent}) starting from ${\cal O}_k$.
An interesting consequence of this rederivation is that we can read the characteristic $E$ in
(\ref{completeness}) from the expression (\ref{confin}):
\be
E=(\vec \Delta, \vec 0).
\label{echar}
\ee
We will provide another check of this identification in the section 6.

To clarify our statement about the existence of the polynomials in (\ref{expansion}), it
is useful to consider in some detail the $SU(2)$ case. The blowup factor is simply given
by \cite{rf:mw}:
\be
\tau(t|u)={\rm e}^{-t^2 T_{2,2} } {\vartheta_4({t \over 2\pi} {du \over da} |\tau) \over
\vartheta_4 (0|\tau)},
\label{ctone}
\ee
where $\vartheta_4$ is the Jacobi theta form with characteristic $[1/2,0]$,
and we put $u \equiv u_2$. The quotient of these theta functions
can be written in terms of a Weierstrass sigma function using the following identity \cite{rf:akh}:
\be
{\vartheta_4 (z|\tau) \over \vartheta_4 (0|\tau)} = {\rm e}^{-\eta_2 \omega_2 z^2} \sigma_3 (\omega_2 z),
\label{sigide}
\ee
where $\omega_2$ is the $a$-period ({\it i.e.}, $\tau=
\omega_1/\omega_2$), and $\eta_2 = \zeta(\omega_2/2)$. Using the
expression for the contact term in (\ref{contnf}), the identity
\be
\eta_2 = {\pi^2 \over 6 \omega_2} E_2 (\tau),
\label{etaid}
\ee
and the fact that $\omega_2=(8 \pi/{\sqrt 2}) (da/du)$, one finds:
\be
\tau (t|u)={\rm e}^{-t^2 {u \over 3}} \sigma_3 ({4 t \over {\sqrt 2}})
\label{onexp}
\ee
This is precisely the expression found in \cite{rf:fs} in the context of Donaldson theory (for
zero magnetic flux). The key fact is that the sigma functions $\sigma_i (t)$ can be expanded around the origin, and the
coefficients of the Taylor expansion are polynomials in the root $e_i$, and the functions $g_2$, $g_3$ (in the Weierstrass
description). These quantities depend only on $u$. From the Seiberg-Witten curve for the
$SU(2)$ pure Yang-Mills theory one finds in fact:
\be
e_3 =-{u \over 12}, \,\,\,\ g_2 = {1 \over 4} \bigl( {u^2 \over 3} - {1 \over 4} \bigr),
\,\,\,\ g_3= {1 \over 48} \bigl( {2 u^3 \over 9} -{u \over 4}\bigr).
\label{exmodu}
\ee
Therefore, the expansion of (\ref{onexp}) is indeed of the form (\ref{expansion}),
and using that
\be
\sigma_3(t)=1 -e_3 t^2 + {\cal O}(t^4),
\label{expsigma}
\ee
it follows that the quadratic term in $t$
in (\ref{onexp}) is zero, as expected from (\ref{conblow}).

If we consider now the higher rank case, the above remarks on $SU(2)$
suggest that the expansion in (\ref{expansion}) should
involve some kind of hyperelliptic generalization of the sigma functions.
It is also quite possible
that the interpretation in terms of the Toda-Whitham
hierarchy gives a constructive way of computing
these polynomials.

\section{Soft Supersymmetry Breaking with Higher  Casimir Operators}
\setcounter{equation}{0}
\indent
In this section, we will give a very rough overview of another use
of Whitham hierarchies: soft supersymmetry breaking with spurion
superfields.

The spurion formalism was introduced in \cite{rf:spurion} and it is a very useful
tool to break supersymmetry in an explicit way. The starting point of the spurion formalism
is a coupling constant in a supersymmetric Lagrangian, call it $m$. This coupling constant
can be promoted to a superfield, $m \rightarrow M$, and this will give another Lagrangian
which will be supersymmetric as well. The superfield $M$ is called a spurion superfield.
Notice that the original Lagrangian is recovered by taking the VEV
of the scalar component of $M$ to be equal to $m$, and setting the rest of the fields
to zero. To break supersymmetry, one gives
a VEV to an {\it auxiliary} field in $M$. The resulting Lagrangian will
be nonsupersymmetric due to the extra terms generated in this way. An interesting example
is the mass term for the ${\cal N}=2$ quark hypermultiplet, which in ${\cal N}=1$
superspace has the form:
\be
\int d^2 \theta m {\widetilde Q} Q,
\label{massnone}
\ee
where ${\widetilde Q}, Q$ are the two ${\cal N}=1$ chiral superfields that correspond
to the ${\cal N}=2$ quark hypermultiplet. If we promote $m$ to an ${\cal N}=1$ chiral superfield:
\be
m \rightarrow M=\phi_m + {\sqrt 2} \theta \psi_m + \theta^2 F_m,
\label{promotion}
\ee
we obtain an ${\cal N}=1$ Lagrangian, and we recover the original one if we put $\langle \phi_m
\rangle =m$. If we want to break supersymmetry, we take:
\be
M=m +\theta^2 F_m.
\label{broken}
\ee
This induces an extra term $F_m {\tilde q} q$ (a mass term for the squark)
which breaks supersymmetry down to ${\cal N}=0$. The VEV of the auxiliary
field, $F_m$, becomes a SUSY breaking parameter. Notice that one can also
take $m=0$, $F_m \not=0$ to generate a mass term for the squarks while keeping
the quarks massless. This is the usual procedure to decouple the squarks
through soft supersymmetry breaking.

We can try to proceed in the same way with ${\cal N}=2$ gauge theories described
by a prepotential. In the most general case, the prepotential is a function
\be
{\cal F} = {\cal F} (\alpha^I, m_f, T_n),
\label{prepo}
\ee
where $T_n$ are the Whitham slow times, and $T_1= \Lambda$. In this Lagrangian,
the variables
$\alpha^I$ correspond to ${\cal N}=2$ vector superfields, while the rest of
the variables are couplings. Any of these couplings can be promoted
to an ${\cal N}=2$ vector superfield, and then one breaks supersymmetry by
giving VEVs to the auxiliary components. This strategy was followed
in \cite{rf:soft,rf:moresoft,rf:luisIyII,rf:hsu} in the context of
the original Seiberg-Witten theory, when $T_n=0$ for $n\ge 2$. The
only coupling constants are then $\Lambda$ and $m_f$.
In the enlarged prepotential, we can construct a more general, nonsupersymmetric
deformation of Seiberg-Witten theory, by promoting $T_{n\ge 2}$ to spurion
superfields while setting the VEVs of their scalar components to zero (for $T_1$, the
scalar component has to be different from zero and equal to the quantum scale
$\Lambda$).
It is convenient to define the couplings $s_n$ as
\be
s_1=-i \log \Lambda, \,\,\,\,\ s_n= -i  T_n, \,\ n=2, \dots, r.
\label{spurions}
\ee
We then promote these couplings to ${\cal N}=2$ vector superfields $S_n$. Such a
superfield can be written, in ${\cal N}=1$ language, as one ${\cal N}=1$ chiral
superfield (that we will also denote by $S_n$) and one ${\cal N}=1$
vector superfield $V_n$.
We then have, in terms of ${\cal N} = 1$ superfields,
\be
S_1 = s_1 + \theta^2 F_1 \;\;\; , \;\;\; V_s \equiv V_1 = \med
D_1 \theta^2 \bar\theta^2 ~,
\label{vevs1}
\ee
\be
S_n = \theta^2 F_n \;\;\; , \;\;\; V_n = \med D_n \theta^2 \bar\theta^2 ~,
\,\,\ n\ge 2.
\label{vevs2}
\ee
The Whitham times are then interpreted as supersymmetry
breaking parameters, and supersymmetry is broken down to ${\cal N}=0$. To obtain the
explicit form of the soft breaking terms, we can expand the prepotential
around $S_{n\ge 2}=0$. The terms of order ${\cal O}(S^3)$
in this expansion do not give any contribution to the Lagrangian, because they involve
too many $\theta$'s and the integral in superspace will vanish. To analyze the softly broken
theory, it is enough to consider the terms which are at most quadratic in the slow times, and the
following expression is exact:
\be
\F (\alpha^I, T_n ) =
\F (\alpha^I, T_1, 0) +
\sum_{n\ge 2} \bigl( {\partial F \over \partial T_n}\bigr)_{T_{n \ge 2} =0} T_n + {1 \over 2}
\sum_{n,m \ge 2}
\bigl( {\partial^2 F \over \partial T_n \partial T_m }\bigr)_{T_{n \ge 2} =0} T_n T_m.
\label{exactex}
\ee
The first term in this expansion is just the Seiberg-Witten
prepotential, after setting $T_1=\Lambda$. Notice that the prepotential defined in the
Whitham framework is defined for the effective theory. Nevertheless, it is
easy to see what is the classical prepotential by
going to the semiclassical region and switching off the quantum corrections.
Again, it is more convenient to
use the reduced prepotential (\ref{reduced}). In this case, the second
derivatives appearing in (\ref{exactex}) vanish semiclassically, as we showed in
(\ref{reder}). The first derivatives can be read from (\ref{lasecudef}). They
are written in terms of Casimir operators and therefore have a well-defined
classical limit. We finally obtain the following expression for the reduced,
classical prepotential:
\be
\F^{\rm red}_{\rm class} = \frac{\beta}{2 \pi} \sum_{n=1}^r \frac{1}{n} S_n \H_{n+1},
\label{linpre}
\ee
where the spurions $S_n$ are given in (\ref{vevs1})--(\ref{vevs2}). The
microscopic Lagrangian associated to (\ref{linpre}) can be
written as follows. First, one defines a generalized matrix of couplings as follows:
\be
\tau_{ab}^{\cl} = {\partial^2 \F^{\rm red}_{\rm class} \over
\partial \phi^a \partial \phi ^b}=\tau\,\delta_{ab} ~,
\label{babcl}
\ee
\be
\tau^{\cl}{_a}{^m} = {\partial^2  \F^{\rm red}_{\rm class} \over
\partial \phi^a \partial S_m }=\frac{N}{\pi{i}m} \pder{\H^{\cl}_{m+1}}{\phi^a} =
\frac{N}{\pi{i}m} \tr\,(\phi^m{T_a}) + \ldots  ~,
\label{bamcl}
\ee
\be
\tau^{mn}_{\cl} = {\partial ^2 \F^{\rm red}_{\rm class} \over
\partial S_m \partial S_n }= 0 ~,
\label{bmncl}
\ee
where the dots in eq.(\ref{bamcl}) denote the derivative with respect
to $\phi^a$ of lower order Casimir operators.
The indices $a,b,c,...$ belong to the adjoint representation of $SU(N)$,
and are raised and lowered with the invariant metric, and $n,m,...$~ correspond
to the variables $s_n$. The complex scalar field of the ${\cal N}=2$ vector multiplet
is written as $\phi=\sum_{a} \phi^a T_a$, where $T_a$ is a basis for the Lie algebra
of $SU(N)$. Finally, $\tau$ is the classical gauge coupling, and it is related to
the spurion $s_1=-i \log \Lambda$ through the one loop formula
$\Lambda^{\beta} = {\rm e}^{\pi i \tau}$. We define now:
\be
b^{\cl} = {1 \over 4 \pi} {\rm Im} \,\tau ^{\cl} ~.
\label{imclass}
\ee
We find
\beqa
{\cal L} & = & {\cal L}_{N=2}  - B^{mn}_{\cl} \left( F_m F^*_{n} +
{1\ov 2} D_m D_n \right) + f^e_{~bc} b^{\cl}{_a}{^m} {b^{\cl}}^{-1}_{ae} D_m
\phi^{b} {\bar\phi}^c
\nonumber \\
& + & {1\ov 8\pi}{\rm Im}
\frac{\partial\tau^{\cl}{_b}{^m}}{\partial\phi^a}\Bigg[
(\psi^a\psi^b)
F^*_m + (\lambda^a\lambda^b) F_m + i\sqrt{2}(\lambda^a\psi^b) D_m\Bigg] ~,
\label{barelag}
\eeqa
where $B^{mn}_{\cl}$ is the classical value of the duality invariant quantity
\be
B^{mn} = b_{a}{^m}{b}^{-1\,ab}b_b{^n} - b^{mn} ~,
\label{moninv}
\ee
and $f^a_{~bc}$ are the
structure constants of the Lie algebra. In (\ref{barelag}), $\lambda$, $\psi$ are the gluinos and
$\phi$ is the scalar component of the ${\cal N} = 2$ vector superfield. We see that the spurion
that corresponds to $s_1$ gives mass to the gauginos of the ${\cal N} = 2$
vector multiplet and to the imaginary part of the Higgs field $\phi$.
The spurions corresponding to higher Casimirs, on the other hand,
give couplings between the Higgs field and the
gauginos. Notice that the spurion
superfields
$S_n$ have dimensions $1-n$, therefore the supersymmetry breaking parameters
$F_n$, $D_n$ have dimension $2-n$. For $n>2$, they will give nonrenormalizable
({\it i.e.}
irrelevant) interactions in the microscopic Lagrangian. This does not mean that
the resulting perturbations do not change the low-energy structure of the
theory: the
operators we are considering can be dangerously irrelevant operators, as in the
related theory analyzed in \cite{rf:seiberg}, and in this case
they will affect the infrared physics.

One of the advantages of this procedure to break supersymmetry is
precisely that one can also write the exact low energy effective
theory associated to (\ref{barelag}), up to two derivatives. This
is due to the fact that the dependence of the Lagrangian on the
SUSY breaking parameters is controlled by the dependence of the
prepotential on the spurions, and this is also given by the
Seiberg-Witten solution. Notice that, in the low energy theory, the
couplings (\ref{babcl})-(\ref{bmncl}) receive quantum corrections,
and can be computed in terms of the hyperelliptic curve data. Using
these couplings, one can write down an effective potential and study
the vacuum structure of the theory. The effect of the
nonsupersymmetric deformations associated to the higher order
Casimirs in the vacuum structure of the theory has been explored in
\cite{rf:our}.

\vspace{1.5cm}

\section{Instanton Corrections}
\setcounter{equation}{0}
\indent
In this section we will briefly review another use of the Whitham hierarchies: the
computation of instanton corrections.
For more details, see \cite{rf:our}, \cite{rf:sanrev} and \cite{rf:jgj}.

One of the main results of the Whitham approach to the theory of the prepotential have been
the equations for the first and second derivatives of the prepotential in (\ref{completeness})
and (\ref{lasecu}) derived in \cite{rf:ITEP}. In \cite{rf:matone} it was realized that the
RG equation
\be
{\partial {\cal F} \over \partial \log \Lambda} ={\beta\ov 4\pi i}  u
\label{mato}
\ee
is very useful to derive a recursion relation for the instanton contributions.
In order to compute the instanton corrections, however, (\ref{mato}) is not
sufficient and
one needs additional input. This is usually provided by the Picard-Fuchs
equations for the
periods. The Picard-Fuchs equations are difficult to derive and solve when the
rank of the gauge group
is larger than one, although techniques from topological Landau-Ginzburg
theories can make them more instrumental in order to obtain the one-instanton
correction to the prepotential for the ADE series \cite{rf:itoyang}. It turns
out that the equation
for $\partial^2 {\cal F}/\partial (\log \, \Lambda)^2$ in (\ref{lasecudef}), together with
(\ref{mato}),
provides enough information to obtain the instanton expansion of the
prepotential in the semiclassical region to any order, and we don't have to
make use of the Picard-Fuchs equations. As we will see, the connection of
$SU(N),~
{\cal N} = 2$ super Yang--Mills theory with Toda--Whitham
hierarchies embodies in a natural way a recursive
procedure to compute all instanton corrections. The essential ingredient that
makes this possible is the relation of the derivatives of the prepotential with
the theta function associated to the root lattice of the gauge group.
To see how this works, consider the
instanton expansion of the prepotential:
\be
\F = {1\ov 2N}\tau_0 \sum_{\vec \alpha_+} Z_{\vec \alpha_+}^2
+ {i\ov 4 \pi} \sum_{\vec \alpha_+} Z_{\vec \alpha_+}^2 \log \, {Z_{\vec \alpha_+}^2\ov
\Lambda^2}
+{1\ov 2 \pi i}\sum_{k=1}^\infty \F_k(Z)
\Lambda^{2Nk}.
\label{elprep}
\ee
In this expression, $\sum_{\vec \alpha_+}$ denotes a sum over
positive roots. The expansion is in powers of $\Lambda^{\beta}$, where
$\beta=2N$ for
$SU(N)$, and $k$ is the instanton
number. We then have
\be
\dpder{\F}{(\log \Lambda)} = {1\ov 2\pi i}
\sum_{k=1}^\infty (2N k)^2\F_k(Z) \Lambda^{2Nk}~,
\label{makdj}
\ee
which, according to (\ref{lasecudef}), should be equal to
\be
\dpder{\F}{(\log\Lambda)\,} =
-{\beta^2\ov 2\pi i} \pder{\H_2}{a^I} \pder{\H_2}{a^J} {1\ov
i\pi}\d_{\tau_{IJ}}
\log \Theta_E(0|\tau).
\label{lateta}
\ee
The derivative of the quadratic Casimir has also an expansion that
can be obtained
from the RG equation and (\ref{elprep}):
\be
\pder{\H_2}{a^I} = {2\pi i\ov \beta} \ppder{\F}{a^I}{\log\Lambda\,}
=C_{IJ} a^J +
 \sum_{k=1}^\infty
 k \F_{k,I} ~ \Lambda^{2Nk}
\label{expanh}
\ee
where $C_{IJ}$ is the Cartan matrix, $\F_{k,I}= \d\F_k/\d a^I$, and we have taken into account that ${1\ov
2N} \sum_{\vec \alpha_+}
Z_{\vec \alpha_+}^2 =
{1\ov 2} a^I C_{IJ} a^J$. The couplings in the semiclassical region are
obtained again from the expansion (\ref{elprep}):
\be
\tau_{IJ}= {i\ov 2\pi}
\sum_{\vec \alpha_+} \pder{Z_{\vec \alpha_+}}{a^I} \pder{Z_{\vec \alpha_+}}{a^J}
 \log \left({ Z^2_{\vec \alpha_+} \ov \Lambda^2} \right)
+{1\ov 2\pi i} \sum_{k=1}^\infty \F_{k,IJ} \Lambda^{2Nk} ~.
\label{latau}
\ee
with $\F_{k,IJ}= \ppder{\F_k}{a^I}{a^J}$. For convenience, in (\ref{latau})  a
term ${i\ov 2\pi}\sum_{\vec \alpha_+} \pder{Z_{\vec \alpha_+}}{a^I}
\pder{Z_{\vec \alpha_+}}{a^J}\left( {2 \pi i \ov N}\tau_0 - 3\right) $ has been
set to zero by a suitable adjustment of the bare coupling $2\pi i\tau_0 = 3N$.
If we set $\vec \alpha= \sum_I n^I\vec \alpha_I$ and define
\be
\Fppk = \sum_{I,J}n^I\F_{k,IJ} n^J,
\ee
we see that the theta function $\Theta_E$ in the
semiclassical region can be written as
\be
\Theta_E(0|\tau) =
\sum_{r=0}^\infty \sum_{\vec \alpha\in\Delta_r} (-1)^{\vec \rho\cdot\vec \alpha}
\prod_{\vec \alpha_+} Z_{\vec \alpha_+}^{-(\vec \alpha\cdot\vec \alpha_+)^2}
\prod_{k=1}^\infty\left(\sum_{m=0}^\infty {1\ov 2^m
m!}\left(\Fppk\right)^m\,
\Lambda^{2Nkm}\right) \Lambda^{2Nr}.
\label{expantheta}
\ee
In the previous expression, $\vec \rho$ is again the Weyl vector, and
$\Delta_r\subset\Delta$ is a subset of the root consisting of the
lattice vectors
$\vec \alpha$ that fulfill the constraint
$\sum_{\vec \alpha_+}(\vec \alpha\cdot\vec \alpha_+)^2
= 2Nr$.
In particular $\Delta_1$ is the root system, {\it i.e.} the simple roots
together with their Weyl reflections. In the above expression, we have used the
characteristic (\ref{echar}), which is the appropriate one if we are working in the electric
frame, as we should do in the semiclassical region. If we insert the expressions (\ref{expanh})
and (\ref{expantheta}) into (\ref{lateta}), and we equate the coefficients of the
different powers of $\Lambda$, we find a set of recursive equations for ${\cal F}_k$
which make possible to compute all the instanton coefficients by starting from
the perturbative contribution to the prepotential.
For example, one finds for the one-instanton correction:
\be
{\cal F}_1  = -
\sum_{\vec \alpha\in\Delta_1}Z_{\vec \alpha}^2~ (-1)^{\vec \rho\cdot\vec \alpha}\prod_{\vec \alpha^+}
Z_{\vec \alpha^+}^{-(\vec \alpha\cdot\vec \alpha^+)^2} ~,
\label{efe1}
\ee
and for the two-instanton contribution:
\beqa
{\cal F}_2 &=& -\frac{1}{4}\left(\sum_{\vec \alpha\in\Delta_1}(-1)^{\vec \rho\cdot\vec \alpha}
\prod_{\vec \alpha_+} Z_{\vec \alpha_+}^{-(\vec \alpha\cdot\vec \alpha_+)^2}
\left[ \F_1 + 2 (\vec \alpha\!\cdot\!\!\F'_k ) Z_{\vec \alpha} + \med (\Fppu ) Z_{\vec \alpha}^2 \right] \right.
\nonumber \\
& & \left. + \sum_{\vec \beta\in\Delta_2}Z_{\vec \beta}^2~ (-1)^{\vec \rho\cdot\vec \beta}\prod_{\vec \alpha^+}
Z_{\vec \alpha^+}^{-(\vec \beta\cdot\vec \alpha^+)^2} \right) ~,
\label{efe2}
\eeqa
where $\vec \alpha\!\cdot\!\!\F'_k = \sum_I n^I\F_{k,I}$.
The above expression makes patent the recursive character of the procedure.
We then see that, with this method, one can find the instanton corrections
to the prepotential in a very simple way. The explicit expressions (\ref{efe1})--(\ref{efe2})
are in full agreement with the results
of \cite{rf:klemm,rf:dhoker,rf:MNW} (see \cite{rf:our} for a detailed presentation).
This agreement gives a further check of the RG
equation (\ref{lasecudef}) and of the choice of the characteristic (\ref{echar}).

These results on the instanton corrections can be extended in many ways. The equation (\ref{lasecudef})
can be also used to study the strong coupling regime
near the points of maximal singularity \cite{rf:jj}. Moreover, one can analyze with this technique
all the classical groups with or without matter content to find general expressions
for the instanton corrections \cite{rf:jgj}.

\section{Conclusions and Outlook}

We have seen that the approach to the ${\cal N}=2$ prepotential based on the
theory of Whitham hierarchies gives a very useful framework to understand
deformations of ${\cal N}=2$ supersymmetric gauge theories. These deformations arise
in different physical contexts. We have explored in detail the structure of the contact
terms in the twisted version of ${\cal N}=2$ theories, and also the nonsupersymmetric
deformations associated to soft supersymmetry breaking. In both cases, the approach
based on Whitham hierarchies provides the right conceptual framework and the technical
tools to derive the precise form of the deformed theory. We have also seen that one can
obtain the instanton expansion in the semiclassical region by using
the equations for the first and second derivatives of the prepotential with respect to
the quantum scale.

Nevertheless, there are many issues that should be
further clarified. The framework of
Whitham hierarchies should be generalized to other gauge groups
and matter content. Some steps in this direction
can be found in \cite{rf:taka2,rf:jgj,rf:next}.
In the context of the twisted theory, one would like
to introduce slow times associated
to any homogeneous combination of the Casimirs,
and construct a prepotential depending on these
slow times and with the right semiclassical behavior.
In this sense, although the explicit
construction given in \cite{rf:ITEP} has
clarified the theory of the prepotential and its
applications to Donaldson theory, one should
be able to improve it along the lines that
we have suggested. It would be also nice to
have an {\it a priori} connection between
the Whitham hierarchy approach to the prepotential,
and the structure of the generating function
for the twisted theory. We have followed a
rather indirect approach, based on a set of constraints
for the contact terms and the RG equations.
The approach of \cite{rf:lns}, based on
Hamiltonian deformations of the prepotential,
should be useful in deriving this
connection.

It would be also very interesting to find an explicit construction
of the polynomials appearing in (\ref{expansion}) for the higher rank case.
As we have suggested, this construction will involve interesting generalizations
of sigma functions for hyperelliptic curves. Another reason for addressing this
issue is that the structure of the blowup formula gives a very direct connection
between the Toda hierarchy underlying the Seiberg-Witten solution, and the
generalizations of Donaldson theory. This relation is certainly intriguing and
suggests that many structures that have been found in two-dimensional topological
gravity and two-dimensional topological matter could be also relevant in four dimensions.

\section*{Acknowledgements}
I would like to thank J.D. Edelstein, J. Mas and G. Moore for the
enjoyable collaborations that led to the results reported here, for
many useful conversations, and for a critical reading of the manuscript.
I would also like to thank the organizers of the workshop ``Gauge theory and
integrable models" for the opportunity to
present these results in a wonderful environment. This work is supported by DOE grant
DE-FG02-92ER40704.

\end{document}